# Solar Spectroscopy at ARIES

K. Sinha[*]

*Aryabhatta Research Institute of Observational Sciences (ARIES), Manora Peak, Nainital - 263 129, India*

**Abstract.**

Identification of Fraunhofer lines with the known atomic and molecular absorbers and predictions leading to such an effort has been a challenging area of study crowned with occasional success. Such studies have also lead, amongst other things to (i) a determination of abundances of elements and that of their isotopes (ii) valuable information on model atmospheres and (iii) use of Sun as a laboratory source. We summarize and review here the work done in the last four decades in the area of solar spectroscopy at Aryabhatta Research Institute of observational sciencES (ARIES in short) with a view to pick up new and interesting areas for future investigations in the light of the tremendous progress made elsewhere in observations of the sun and in the laboratory studies.

*Keywords* :
Solar Spectroscopy, Solar Atlas, Molecules, Model Atmospheres, Line Identifications.

## 1. Introduction

Studies on solar spectroscopy have been an important area of activity at ARIES. Recently, Melendez and Barbuy (1999) extended our work to include several hundred additional lines for a determination of log gf values in the J and H bands with the use of the solar spectrum. Also, based on our earlier predictions, Berdyugina and Livingston (2002) reported detection of the mercapto radical SH in the solar photosphere. Though large equivalent widths (EWs in short) are predicted for these lines in the umbral spectrum, in view of strong scattered light in UV from the surrounding photosphere, the results of a search by these authors in sunspots were unsatisfactory. Further, amongst the more recent work in umbral spectroscopy the work by Sonnabend et al (2006) can be cited. We feel prompted to summarize our work in the following with a view to identify and pick up new areas for studies.

---

[*]e-mail:ksinha2000@hotmail.com

The work at ARIES has been more focused on theoretical investigations on molecules in sunspots, photosphere and faculae with a few papers on atomic lines and where possible to compare the results with published observational data. We also tried to develop without much success a 25 cm horizontal f/66, $6^0$ off-axis skew cassegrain telescope together with a high dispersion double pass spectrograph. Work in the then contemporary scene has been summarized in reviews by Pande (1972) and Sinha (1991, 1993, 1998).

In the following, it is not our intention to underplay or to take away any credit from colleagues who independently worked elsewhere on similar lines. In fact, we received lot of help, support and encouragement from these colleagues who worked in the frontier areas of research.

## 2. Predictions on line identifications

The molecules AlO, BO, CaO, $CH_2$, $CO_2$, CS, HCl, HCN, HCO, $H_2$, $H_2S$, NaH, $NH_2$, $NH_3$, NO, NS, $N_2$, $O_2$, PH, PN, PO, SH, SiO, SO, $S_2$, $TiO_2$, VO, YO and $AlH^+$, $CO^+$, $H_2^+$, $MgH^+$, $NH^+$, $OH^+$, $SH^+$, $SiH^+$ were considered important in sunspots on the basis of number/partial pressure considerations. EW calculations showed that the vibration - rotation lines of HCl, HF, MgH, NO, SiH and SiO and UV transitions in $MgH^+$, NO, $O_2$, PH, PO, SH and SiO were expected present and also that the isotopic lines of SiO and SH could additionally be looked for in the umbral spectrum. The molecules BO, HI, HBr and HCN were found not detectable. SiO lines were detected in the umbral spectrum by Glenar et al (1983).

Dissociation equilibrium calculations showed that the molecules CS, HCl, HF, SH, SiO and SO form in sufficient numbers and that they could show up in the corresponding photospheric spectrum. Further, the concentrations of the considered ionic species dropped in the order $H_2^+$, $CO^+$, $OH^+$, $NO^+$, $O_2^+$ and $N_2^+$. EW calculations indicated that observable but very weak lines of CS, $C_2$ (Phillips, Ballik - Ramsay and Mulliken bands), $MgH^+$ and SH were also there in the photospheric spectrum. Some of these predictions were later commented upon or supported with observations by Lambert and Mallia (1974), Brault et al (1982) and Grevesse et al (1977). We identified additional lines of the $C_2$ Swan system in the photospheric spectrum. $CO^+$, $HeH^+$ and NO - gamma system were found absent and we could not convince ourselves with the detection of $CH^+$ in photosphere.

We found that in the facular spectra, $CH^+$ could show up whereas $MgH^+$, $OH^+$ and $SiH^+$ could also be present but $AlH^+$, $CN^+$, $CO^+$, $N_2^+$ and $NH^+$ were not detectable.

## 3. Sensitivity to Model Atmospheres

Using TiO lines of the alpha system, rotational temperature, $T_{rot}$, consistent with Zwaan's sunspot model was obtained. Also, center to limb (CL in short) behavior of Trot for the infrared bands of SiO in different sunspot models was theoretically investigated and the same was found to vary from 150 to 400 K from model to model.

For the MgH green band lines in sunspots, an inequality $W_{cal} >> W_{obs}$ was found and to resolve the discrepancy, the role of penumbral blurring and scattered light from umbral dots was stressed upon. We also suggested a revision of the $\Delta v \neq 0$ oscillator strengths.

The concentrations of molecules was shown to vary with magnetic field strengths and EW of a representative CO line was found to linearly increase initially and attain a saturation value with increase in magnetic field strength. The simultaneous observations in photosphere and spots for $C_2$, MgH and TiO found in a narrow spectral range was theoretically demonstrated a good probe to sense the evolution of sunspots with increasing diameters and also to sense if the sunspots are different at different phases of the solar cycle.

Our own observations on photospheric CH lines at different CL distances for $T_{rot}$, do not help decipher a change in the values in view of a large scatter. Fresh observations as well as the use of Delbouille et al's (1973) atlas for the CN and CH lines for a determination of doppler widths has shown that the same varies linearly with EW. Also, a variation in the turbulence velocities towards the limb was inferred.

Observations at the center of the solar disc for the photospheric $C_2$ and MgH were found not to reproduce the model based rotational temperatures. The problem was further investigated in detail and we utilized the published EWs at different CL distances from Withbroe (1968) but a spurious rise in $T_{rot}$ at a near limb position was obtained. The discrepancy was resolved by accounting for saturation in lines which is very important for $C_2$. However, we still faced a small problem; for $\mu = 0.2$ on the solar disc, against a model based $T_{rot} = 5110 \pm 10$, an observed value $T_{rot} = 5400 \pm 240$ was obtained by us. It lead us to point out that the temperature in the outer layers of the photospheric model due to Holweger and Mueller (1974) needed slight increases. The task was later accomplished independently by Grevesse and Sauval (1999) to solve the controversy surrounding iron abundances.

The behavior of a few lines of the Balmer, Paschen and Brackett series of hydrogen and their CL variation in representative facular and photospheric models were investigated with the conclusion that these lines were sensitive probes for the chosen model atmospheres. Similar conclusions on the use of the vibration rotation lines of CO and the red and the violet system lines of CN were reached. EWs for the CH, $C_2$, MgH, NH, OH and SH lines were calculated for future comparison with observations and it was found that CH, NH and OH could be good candidates for testing model atmospheres. Rotational temperatures, $T_{rot}$, for the vibration rotation lines of CO were reported to be consistent with the chosen facular models.

## 4.  Sun as a laboratory source

Utilizing the published material on solar spectrum, the oscillator strengths of the Phillips bands of $C_2$, the green bands of MgH and that of SiH$^+$ were derived. The dissociation energy, $D_0{}^0$, for the CN molecules was found to lie in the range 7.66 to 7.76 eV (Sinha and Tripathi, 1986) which is supported by even recent investigations (Reddy et al, 2003; Gustafsson and Wahlin, 2006). Log gf values of Fe, Fe$^+$, Ni, Sc, Si, Ti and V in the spectral range 6209 - 6273 Å were derived and so also for about 300 lines due to the 17 different atomic species in the J and H bands.

Franck - Condon factors for UV transitions in AlF, CaH, $P_2$, SiCl, SiF and $S_2$ and also the partition functions and the dissociation constants for HeH$^+$ and MgH were calculated. For the $H_3{}^+$ ion only partition functions were calculated. We made suggestions on life time and wave number measurements of the infrared transitions in MgO. The wave numbers have since become available (Kagi et al, 1994). Rotational and vibrational constants for CaH$^+$ are theoretically calculated and reported.

## 5.  Conclusions

Thus, it is seen that fresh attempts towards identification of spectral lines, towards refining a model atmosphere and towards a study of evolution of sunspots with the help of molecular lines hold promise for future studies. A beginning in this direction can be made with the available FTS solar atlases. In the infrared, vibration - rotation transitions in MgH, NO and SiH and in the ultra - violet, the electronic transitions in MgH$^+$, NO, $O_2$, PH, PO, SH and SiO could be important for identifications in the solar spectrum.

A bibliography on solar spectroscopy at ARIES is available at http://aries.ernet.in/Research/Bibliography.pdf.